\documentclass[twocolumn,floats,floatfix,amssymb,nofootinbib,prd,superscriptaddress]{revtex4-2}
\usepackage{graphicx}
\usepackage{amsmath,amssymb}
\usepackage{amsfonts}
\usepackage{xspace} 
\usepackage[usenames]{color}
\usepackage{dcolumn}
\usepackage{bm}
\usepackage{mathrsfs}
\usepackage[colorlinks=true]{hyperref}
\usepackage[all]{hypcap} 
\usepackage[utf8]{inputenc} 
\usepackage{multirow}
\usepackage{etoolbox}
\usepackage{tikz}

\def\be{\begin{equation}}
\def\ee{\end{equation}}
\def\bea{\begin{eqnarray}}
\def\eea{\end{eqnarray}}

\newcommand{\beq}{\begin{eqnarray}}
\newcommand{\eeq}{\end{eqnarray}}

\newcommand{\bit}{\begin{itemize}}
\newcommand{\eit}{\end{itemize}}
\newcommand{\ben}{\begin{enumerate}}
\newcommand{\een}{\end{enumerate}}

\newcommand{\tn}{\textnormal}

\begin{document}

\title{Gravitational waves from extreme-mass-ratio systems \\in astrophysical environments}

\author{Vitor Cardoso}
\affiliation{Niels Bohr International Academy, Niels Bohr Institute, Blegdamsvej 17, 2100 Copenhagen, Denmark}
\affiliation{CENTRA, Departamento de F\'{\i}sica, Instituto Superior T\'ecnico -- IST, Universidade de Lisboa -- UL,
Avenida Rovisco Pais 1, 1049 Lisboa, Portugal}
\author{Kyriakos Destounis}
\affiliation{Dipartimento di Fisica, Sapienza Università di Roma, Piazzale Aldo Moro 5, 00185, Roma, Italy}
\affiliation{INFN, Sezione di Roma, Piazzale Aldo Moro 2, 00185, Roma, Italy}
\affiliation{Theoretical Astrophysics, IAAT, University of T{\"u}bingen, 72076 T{\"u}bingen, Germany}
\author{Francisco Duque}
\affiliation{CENTRA, Departamento de F\'{\i}sica, Instituto Superior T\'ecnico -- IST, Universidade de Lisboa -- UL,
Avenida Rovisco Pais 1, 1049 Lisboa, Portugal}
\author{Rodrigo Panosso Macedo}
\affiliation{STAG Research Centre, University of Southampton, University Road SO17 1BJ, Southampton, UK}
\author{Andrea Maselli}
\affiliation{Gran Sasso Science Institute (GSSI), I-67100 L’Aquila, Italy}
\affiliation{INFN, Laboratori Nazionali del Gran Sasso, I-67100 Assergi, Italy} 
\begin{abstract} 
We establish a generic, fully-relativistic formalism to study gravitational-wave emission by extreme-mass-ratio systems in spherically-symmetric, non-vacuum black-hole spacetimes. The potential applications to astrophysical setups range from black holes accreting baryonic matter to those within axionic clouds and dark matter environments, allowing to assess the impact of the galactic potential, of accretion, gravitational drag and halo feedback on the generation and propagation of gravitational waves. We apply our methods to a black hole within a halo of matter. We find fluid modes imparted to the gravitational-wave signal (a clear evidence of the black-hole fundamental mode instability) and the tantalizing possibility to infer galactic properties from gravitational-wave measurements by sensitive, low-frequency detectors.
\end{abstract}

\maketitle 

\noindent{\bf{\em Introduction.}}
The birth of gravitational-wave (GW) astronomy ushered in a new era in gravitational physics and high-energy astrophysical phenomena \cite{LIGOScientific:2016aoc,LIGOScientific:2018mvr}. GWs carry unique information about compact objects, most notably black hole (BH) systems, and grant us access to exquisite tests of the gravitational interaction in the strong field, highly dynamical regime \cite{Yagi:2016jml,Barack:2018yly,Cardoso:2019rvt,Baibhav:2019rsa,LIGOScientific:2021sio,Cheung:2022rbm,Mitman:2022qdl}. 

They also bear precious information about the environment where compact binaries live~\cite{Yunes:2011ws,Barausse:2014tra,Derdzinski:2020wlw,Cardoso:2020iji,Zwick:2022dih}. This knowledge is important {\it per se}, and may inform us on how compact binaries are formed~\cite{Pan:2021oob} or how BHs grow and evolve over cosmic times~\cite{Cardoso:2022nzc}. In addition, GWs are sensitive to accretion disk properties~\cite{Speri:2022upm} and even on fundamental aspects, such as the existence of dark matter spikes in galactic centers~\cite{Eda:2013gg,Macedo:2013qea,Kavanagh:2020cfn,Speeney:2022ryg}; on possibly new fundamental degrees of freedom that can condense around spinning BHs~\cite{Brito:2015oca,Maselli:2021men}; and finally on the nature and existence of BHs, as well as whether they are well described by the Kerr family, a quest which demands environmental effects to be disentangled from purely gravitational ones.

The above questions require a precise modeling of compact binaries in a fully-relativistic setting. Unfortunately, the state-of-the-art adopts at least one of the following approximations: a slow-motion quadrupole formula to estimate GW emission and the dynamics \cite{Babak:2006uv,Destounis:2020kss,Destounis:2021mqv,Destounis:2021rko}, Newtonian dynamical friction, or considers vacuum backgrounds. Recent attempts to refine the analysis by including some relativistic effects indicate that these can have a significant impact on the conclusions one makes regarding detectability and parameter estimation \cite{Speeney:2022ryg,Vicente:2022ivh,Traykova:2021dua}.

Here -- based on classical works on perturbation theory~\cite{1967ApJ...149..591T,Detweiler:1985zz,Chandrasekhar:1991fi,Kojima:1992ie,Allen:1997xj,Sarbach:2001qq,Martel:2005ir,Martel:2003jj} -- we develop a generic, fully-relativistic formalism to handle environmental effects in extreme-mass-ratio inspirals (EMRIs) in spherically-symmetric, but otherwise generic, backgrounds. These are inherently relativistic systems, expected to populate galactic centers and be observable with the upcoming space-based LISA mission \cite{Gair:2017ynp,Pan:2021ksp,Amaro-Seoane:2022rxf,LISA:2022kgy}, and for which Newtonian approximations are ill-suited. Our framework is able to treat GW generation and propagation, but also includes matter perturbations and therefore is able to capture other environmental effects, such as dynamical friction~\cite{Traykova:2021dua, Vicente:2022ivh}, accretion and halo feedback, and will be important to understand mode excitation or depletion of accretion disks, and even viscous heating in these systems. We use geometric units $G=c=1$ everywhere.

\noindent{\bf{\em Setup.}}
We wish to study a static, spherically-symmetric spacetime describing a BH immersed in some environment, like an accretion disk or a dark matter halo, with line element,
\begin{equation}
ds^2=g_{\mu\nu}^{(0)}dx^\mu dx^\nu=-a(r)\,dt^2+\frac{dr^2}{b(r)}+r^2d\Omega^2\, , \label{math:backgroundmetric}
\end{equation}
where $d\Omega^2$ is the line element of the 2-sphere, and characterized by a (anisotropic) stress tensor~\cite{Raposo:2018rjn} 
\beq
T_{\mu\nu}^\tn{env(0)}= \rho u_\mu u_\nu + p_r k_\mu k_\nu + p_t \Pi_{\mu\nu} \, ,
\eeq
where $\rho$ is the total energy density of the fluid, $p_r$ and $p_t$ are its radial and tangential pressure respectively, $u^\mu$ the 4-velocity of the fluid, $k^\mu$ a unit spacelike vector orthogonal to $u^\mu$, such that $k^\mu k_\mu = 1$ and $u^\mu k_\mu = 0$, and $\Pi_{\mu\nu}= g_{\mu\nu} + u_\mu u_\nu - k_\mu k_\nu$ is a projection operator orthogonal to $u^\mu$ and $k^\mu$ (environmental quantities are hereafter denoted with a superscript ``env''). The functions $a(r)$ and $b(r)$ are to be determined by the physics; to prevent clustering throughout the text we drop the $(t,r)$ dependence from all functions, unless necessary. We leave them general for most of the main body, but specialize to the physics of a supermassive BH surrounded by a halo of matter when necessary. The corresponding solution, which we will term galactic BHs (GBHs), was recently derived \cite{Cardoso:2021wlq} and is characterized by the BH mass $M_{\rm BH}$, halo mass $M$ and its spatial scale $a_0$ (see also \cite{Konoplya:2022hbl,Destounis:2022obl} for generalizations and applications).

We now envision a secondary object of mass $m_p$ (a star, asteroid or stellar-mass BH for example) orbiting the above primary BH and causing perturbations to the geometry and matter stress tensor,
\beq
g_{\mu\nu}=g^{(0)}_{\mu\nu}+ g^{(1)}_{\mu\nu},\,\,\,
T^\tn{env}_{\mu\nu}&=&T^\tn{env(0)}_{\mu\nu}+T^\tn{env(1)}_{\mu\nu},
\eeq
where a superscript ``(1)'' denotes perturbations.

The spherical symmetry of the background allows for a separation of variables in the first-order quantities, expanding into tensor spherical harmonics, classified as {\it axial} and {\it polar}, according to their properties 
under parity~\cite{Regge:1957td,Zerilli:1970se,Zerilli:1970wzz}. In the Regge-Wheeler gauge \cite{Regge:1957td,Zerilli:1970se,Zerilli:1970wzz,Sarbach:2001qq,Martel:2005ir}, these are defined by radial functions $h_{0}^{\ell m },\,h_{1}^{\ell m }$ (axial) and $K^{\ell m }, H_{0}^{\ell m }, H_{1}^{\ell m }, H_{2}^{\ell m }$ (polar), and a set of angular basis functions~\cite{Sago:2002fe,Sarbach:2001qq,Martel:2005ir}.

The perturbations induced by the orbiting object on the environment are known once
its pressure, density and velocity fluctuations are computed. These can also be expanded in harmonics. For example, a scalar quantity $X=p_t,p_r,\rho$ will have a perturbation $X^{(1)}$ expanded as
\be
X^{(1)}=\sum_{\ell=2}^\infty\sum_{m=-\ell}^\ell\delta X^{\ell m}(t,r)Y^{\ell m}(\theta,\phi)\label{math:rhopert}\ ,
\ee
with $Y^{\ell m}(\theta,\phi)$ being the standard spherical harmonics on the two-sphere. A similar procedure is applied to any vector quantity.

Finally, a barotropic equation of state provides a further relation between pressure, density variations and the medium's speed of sound via
\beq
\delta p_{t,r}^{\ell m}(t,r)=c_{s_{t,r}}^2(r) \, \delta \rho^{\ell m}(t,r)\,.
\eeq
Here, $c_{s_{r}}(r)$ and $c_{s_{t}}(r)$ are, respectively, the radial and transverse sound speeds. The explicit perturbed equations are shown in the \textit{Supplemental Material}  (see also Ref.~\cite{Kojima:1992ie} if $a=b$). 

With the above procedure, perturbations to the environmental stress-tensor are completely characterized. The source of these perturbations is modeled as a pointlike object with stress tensor
\be
T^{\mu\nu}_p = m_p \int u^\mu_p u^\nu_p \frac{\delta^{(4)}\left(x^\mu - x_p^\mu(\tau) \right)}{\sqrt{-g}} d\tau \, , 
\label{eq:StressEnergyPP}
\ee
where $m_p$ is the mass of the secondary, $\tau$ its proper time, $x^\mu_p(\tau)$ its world-line and $u^\mu_p = d x^\mu_p/d\tau$ its 4-velocity. This stress-energy tensor can also be decomposed in terms of the angular basis \cite{Zerilli:1970wzz,Sago:2002fe}, thereby separating the equations of motion. We will always assume that the pointlike secondary is on a geodesic of the background spacetime~\eqref{math:backgroundmetric}, and use this to simplify the equations of motion. 

\noindent{\bf{\em Evolution equations.}}
%
The perturbations are described by wave equations with a principal part expressed in terms of the operator
$
{\cal L}_v= v^2 \partial^2 /\partial r_*^2 -\partial^2/\partial t^2,
$
with $v$ the field's characteristic speed of propagation. Specifically, axial perturbations propagate with the speed of light $v=1$ and are simply described in terms of a master variable
$\chi=h_1^{\ell m} \sqrt{a b}/r $, governed by the equation
\beq
&&{\cal L}_1\chi- V^{\rm ax}\chi = S^{\rm ax}\,,\label{eq:master_axial} \\
&&V^{\rm ax}=\frac{a}{r^2}\left(\ell(\ell+1) - \dfrac{6m(r)}{r} + m'(r) \right)\,,
\eeq
with $m(r) = r \left(1 - b\left(r\right)\right) /2 $,
the tortoise coordinate is defined by $dr_*/dr=\sqrt{ab}$, and the source term depends on the motion of the point particle (explicit expressions for circular motion are shown in the \textit{Supplemental Material}). The polar sector can be re-expressed as a system of 3 ``wavelike'' equations for $\vec{\phi}=(S,K,\delta\rho)$ 
\be
\hat {\cal L} \vec{\phi}= \hat {\textbf{B}} \vec{\phi}_{,r^*} + \hat{\textbf{A}}\vec{\phi}+\vec{S}_1\,,
\ee
with $S = a/r \,( H_0 - K) $, and $\hat{\cal L} \vec{\phi}=\left({\cal L}_1\phi_1,{\cal L}_1\phi_2,{\cal L}_{c_{s_r}}\phi_3\right)$, i.e., $\phi_1,\,\phi_2$ have characteristic velocity $v=1$, and $\phi_3$ has $v=c_{s_r}$.

We also study perturbations in the frequency domain by Fourier-transforming the evolution equations. Instead of a second-order system for the polar sector, we worked instead with the first-order system
\begin{equation}
\frac{d\vec{\psi}}{dr}=\hat{\boldsymbol{\alpha}}\vec{\psi}+\vec{S}_2\ ,
\end{equation}
with $\vec{\psi}=(H_1,H_0,K,W,\delta\rho)$, and $W$ a fluid velocity quantity.
The matrices $\hat{\textbf{A}},\hat{\textbf{B}},\,\hat{\boldsymbol{\alpha}}$, as well as source vectors $\vec{S}_i$ are shown in the \textit{Supplemental Material}.
Particle contributions enter as a source term $\vec{S}_1,\,\vec{S}_2$ for the metric variables. 

We solve the above problem with two independent codes, based on different approaches, one in the time and the other in the frequency domain. Both use a smoothed distribution to approximate the point particle, $\sqrt{2\pi}\sigma\delta(r-r_p)=\exp{\left(-(r-r_p)^2/(2\sigma^2)\right)}$ where the width $\sigma$ is varied to assess numerical convergence. In the axial sector, the time domain code follows Ref.~\cite{Zenginoglu:2011zz, Sundararajan:2007jg} which places the outer boundary condition at future null infinity by using the same hyperboloidal layers employed there. In the polar sector, the equations are solved in the usual radial tortoise cooordinate with physical  boundaries placed sufficiently far, so that the physical quantities are extracted within the wave equation's causality domain and in a near vacuum region. For example, if we evolve the system for $t=10^3M_\text{BH}$ and extract at $r_*^{\rm ext} = 500M_\text{BH}$, then the outer boundary should be placed further than $r_*^{\rm out} = 10^3M_\text{BH}$ to prevent any signal from being reflected back and affect the field values at the extraction radius. Unless stated otherwise, we use $r_*^{\rm ext}=\text{max}\{10^2a_0,10^3 M_\text{BH}\}$ as extraction radius in the time domain code for the polar sector. The frequency domain code follows the framework from Refs.~\cite{Cardoso:2016olt} in both sectors, with outer physical boundaries placed at $r^{\rm ext}=\text{max}\left\{10^3 / \Omega_p, 2a_0 \right\}$, with $\Omega_p$ the orbital angular frequency. For the gravitational perturbations, we impose usual outgoing boundary conditions there and vanishing Dirichlet boundary conditions for the matter variables. The results from all codes agree within the numerical error when varying these parameters.
Once the metric variables are computed, fluxes in GWs can be calculated. Our two codes are made freely available to the community~\cite{GRIT_REPO,SGREP_REPO}.

\noindent{\bf{\em Boundary conditions and sound speed.}}
%
\begin{figure}[t]
    \centering
    \includegraphics[width=\columnwidth]{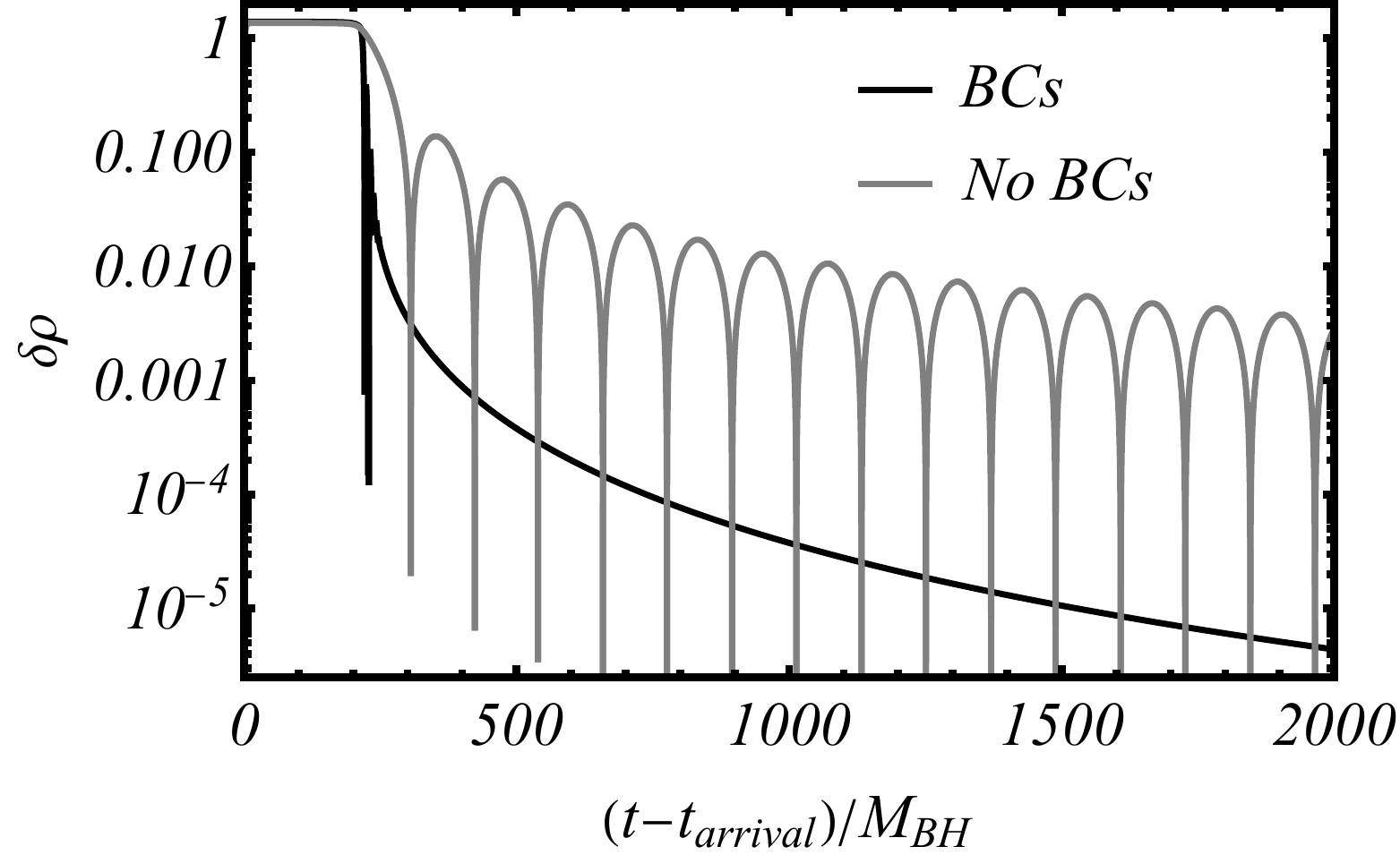}
    \caption{Evolution of $\delta \rho$ in a Schwarzschild background with $c_{s_r}=0.9, c_{s_t}=0$ with different boundary conditions imposed. $t_\text{\textit{arrival}}$ corresponds to the time of arrival of the first direct signal. When $\delta \rho$ is left free at the horizon, an oscillatory tail sets in at late times, consistent with that of a scalar field of mass $\mu_{\rm eff}\,c_{s_r}$. Instead, when Dirichlet conditions are imposed at some cutoff radius $r_\text{cut}$ (here $r_\text{cut}=3M_\text{BH}$), we find a universal power-law decay independent of $r_\text{cut}$ and $c_{s_r}$.}
    \label{fig:deltarho_inf}
\end{figure}
Environments cause the presence of density waves that couple to gravity.
To understand their asymptotic behavior, it's sufficient to examine a vacuum BH background of mass $M_{\rm BH}$, to which the field equations reduce very far or very close to the horizon. For {\it constant} sound speeds, with the ansatz $\delta\rho=r^{\alpha}(r-2M_{\rm BH})^\beta \Psi$, we find that $\Psi$ is governed by the wave equation ${\cal L}_{c_{s_r}}\Psi -{\rm V} \Psi=0$ for
\be
\alpha=\frac{1}{4}\left(-5+\frac{1+4c_{s_t}^2}{c_{s_r}^2}\right)\,,\quad\beta=-\frac{3}{4}-\frac{1}{4c_{s_r}^2}\,,
\ee
with ${\rm V}={\cal O}(r^{-2})$ at infinity and
${\rm V}=\left(\frac{1-c_{s_r}^2}{8c_{s_r}^2M_{\rm BH}}\right)^2$ at the horizon.
The explicit form of ${\rm V}$ and wave equation for $\Psi$ are identical to that obtained in Ref.~\cite{Allen:1997xj} for isotropic fluids, with a suitable change of wavefunction $H$, once we identify $c_{s_r}=c_{s_t}$. Thus, close to the horizon density fluctuations propagate as an effectively massive scalar of mass $\mu_{\rm eff}=\frac{1-c_{s_r}^2}{8c_{s_r}^2M_{\rm BH}}$. A rigorous analysis of the wave equation above is required to understand all the details of the density waves around BHs; however, based on knowledge of massive fields around BHs \cite{Ching:1995tj,Hod:1998ra,Koyama:2001ee}, we expect an intermediate-time power-law tail of the form $\Psi\sim t^{-5/6}\sin\left( \mu_{\rm eff}\,c_{s_r}\right)$, caused by back-scattering in the near-horizon region and probably giving way to another power-law behavior dictated by the asymptotic region far from the BH~\cite{Koyama:2001ee}. Our numerical results in Fig. \ref{fig:deltarho_inf} -- for initial conditions
$\delta \rho = 0 \,,\partial_t \delta \rho = \exp(-(r_* - 100M_{\rm BH})^2 / 2)$, extracted at $r_* = 1000M_\text{BH}$ -- support this claim. We find excellent agreement with an oscillatory term $\sin\left( \mu_{\rm eff}\,c_{s_r}\right)$ and decay $t^{-5/6}$. We find a similar behavior for other values of $c_{s_r}$. 

Configurations with a matter profile that vanishes at the horizon and spatial infinity, have sound speeds expected to vanish asymptotically. For sound speed profiles that vanish as a power-law at the boundaries, we find that regular density fluctuations $\delta\rho$ must satisfy Dirichlet conditions. We implement this restriction keeping $c_{s_r}$ constant everywhere, but imposing Dirichlet conditions on fluid variables at some cutoff radius $r_\text{cut}$ close to the BH. 
It is now possible to prove that the late time asymptotics is governed not by the near-horizon but by the large-$r$ asymptotic behavior and that the field should decrease as $t^{-3}$, {\it independently of the multipole $\ell$} \cite{Ching:1995tj}. This is seen clearly in our simulations in Fig. \ref{fig:deltarho_inf}. The direct signal is followed by a universal power-law tail $\delta \rho \sim t^{-3}$, independently of cutoff radius $r_\text{cut}$ and sound speed $c_{s_r}$. 

\noindent{\bf{\em Environment and spectral stability.}}
%
\begin{figure}[t]
    \centering
    \includegraphics[width=\columnwidth]{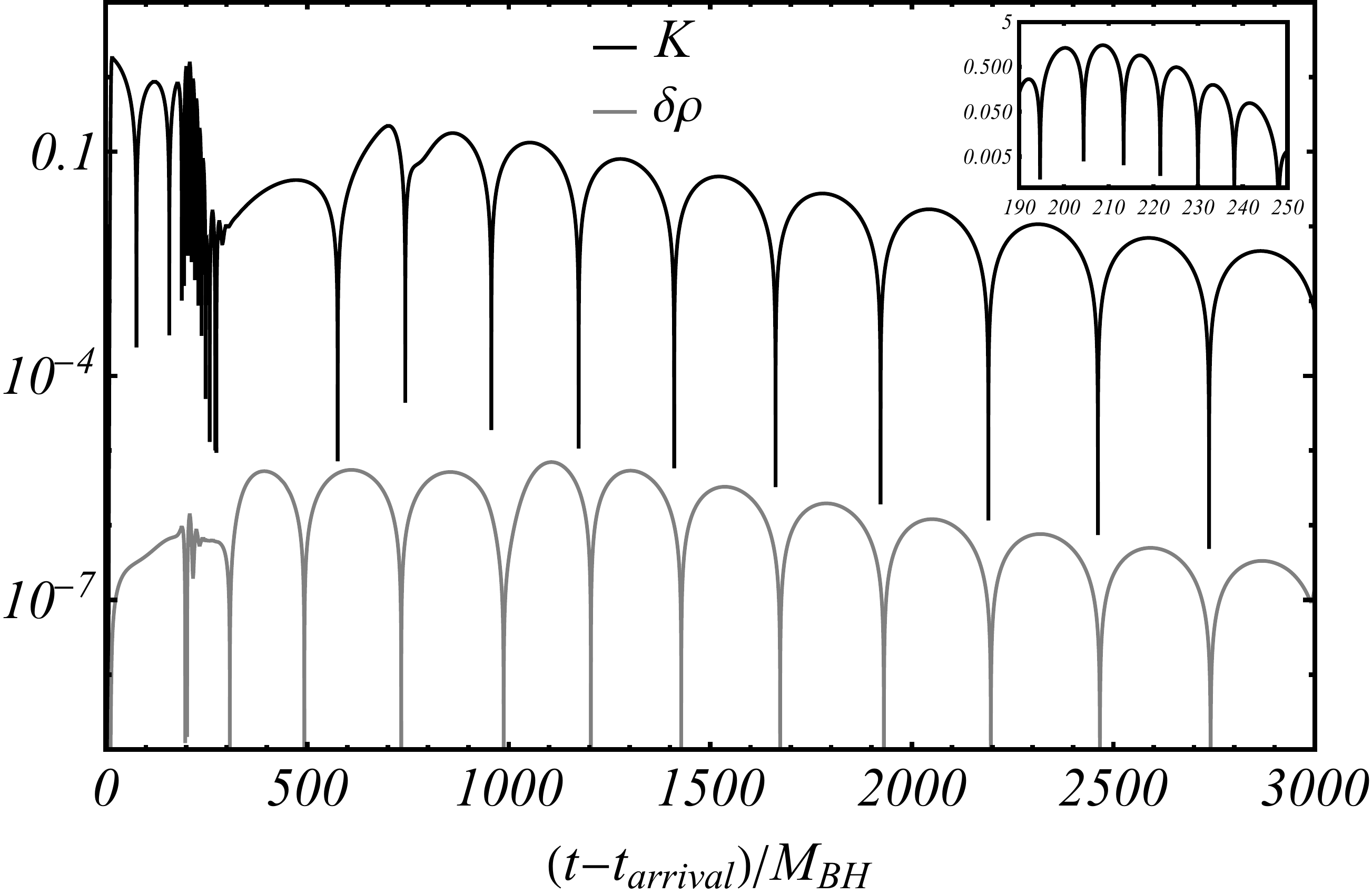}
    \caption{Evolution of the metric and density perturbation $K,\delta\rho$, with $M=10M_\text{BH}$, $a_0 = 10M$. We impose Dirichlet conditions at $r_{\rm cut}=3M_\text{BH}$ and $c_{s_r}=\left[\left(2M_{\rm BH} + a_0\right)/\left(r + a_0\right)\right]^4$, so that it asymptotes to zero at large distances. At early times, BH ringdown is excited (inset for $K$); at late times, we observe a slowly-decaying, fluid-driven mode with period $\propto a_0$. Notice a mutual conversion between GWs and density waves.}
    \label{fig:K_halomodes}
\end{figure}
From now on, we always work with vanishing sound speeds at the boundaries. It is clear from the above that there are two characteristic speeds in the problem, the radial sound speed $c_{s_r}$ and the light speed. Accordingly, and because the polar sector is coupled, we expect to have two families of perturbations, one led by gravity, traveling at the speed of light, the other led by matter fluctuations, traveling at $c_{s_r}$. A clear example of the importance of this coupling is seen through scattering a gaussian wavepacket of {\it gravitational} waves (initial conditions identical to those of Fig.~\ref{fig:deltarho_inf}, but for the metric function $K$). The metric perturbation $K$ and $\delta \rho$ are shown in Fig.~\ref{fig:K_halomodes}. We see conversion from GWs to density waves and vice-versa; BH ringdown at early times, and a long-lived mode at late times. This is in essence a fluid mode, imprinted on the GW signal due to the coupling, and a clear example of spectral instability in BH quasinormal modes, which has attracted considerable interest recently \cite{Barausse:2014tra,Jaramillo:2020tuu,Jaramillo:2021tmt,Destounis:2021lum,Cheung:2021bol,Berti:2022xfj}, here seen in a realistic astrophysical setting.

\noindent{\bf{\em Fluxes from orbiting particles.}}
%
\begin{table}[ht!]
	\begin{tabular}{c c c c c c} 
		\hline\hline
		$\ell$ & $m$ & $\dot{E}^t_\infty$ & $\dot{E}^f_\infty$ & $\dot{E}^{\rm BHPT}_\infty$ \\
		\hline\hline
		\multirow{2}*{2} & \multirow{2}*{1} & 8.1629e-7 & 8.1631e-7 & 8.1631e-7 \\ 
						 &	  			    & 6.9156e-7 & 6.9158e-7 &		    \\  
		\hline
		\multirow{2}*{2} & \multirow{2}*{2} & 1.7068e-4 & 1.7062e-4 & 1.7062e-4 \\     			   &	  			  & 1.6077e-4 & 1.6208e-4 &			  \\
		\hline  
		\multirow{2}*{3} & \multirow{2}*{2} & 2.5198e-7 & 2.5199e-7 & 2.5198e-7 \\
					     &	  			    & 2.1611e-7 & 2.1612e-7 &			\\
		\hline
		\multirow{2}*{3} & \multirow{2}*{3} & 2.5490e-5 & 2.5473e-5 & 2.5471e-5 \\
						 &	  			    & 2.3163e-5 & 2.3140e-5 &			\\
		\hline
		\multirow{2}*{4} & \multirow{2}*{3} & 5.7750e-8 & 5.7749e-8 & 5.7749e-8	\\
						 &	  			    & 5.0252e-8 & 5.0252e-8 &			\\
		\hline
		\multirow{2}*{4} & \multirow{2}*{4} & 4.7352e-6 & 4.7260e-6 & 4.7253e-6	\\
						 &	  			    & 4.0458e-6 & 4.0823e-6 &	        \\
		\hline\hline
	\end{tabular}
\caption{Energy flux (in units of $m_p^2/M_\text{BH}^2$) emitted to infinity in different modes by a particle in circular orbit around a GBH at radius $r_p = 7.9456 M_{BH}$. We show results for vacuum (first line of each mode) and for GBH with $c_{s_{r,t}}=(0.9,0)$, $M=10M_\tn{BH}$ and $a_0=10M$. $\dot{E}^t_\infty$ is computed with a time domain integrator, $\dot{E}^f_\infty$ in the frequency domain and $\dot{E}^{\rm BHPT}_\infty$corresponds to results from the BHPT, available only in vacuum. $\ell=m$ modes correspond to polar excitations whereas $\ell=m+1$ correspond to axial ones.
}\label{tab:Fluxes_Comparison}
\end{table}
We have tested our procedure and routines in the vacuum limit, i.e. using a GBH geometry~\cite{Cardoso:2021wlq} with low value of the halo mass $M=10^{-6}M_\tn{BH}$, comparing the GW fluxes with those obtained by the Black Hole Perturbation Toolkit (BHPT) \cite{BHPToolkit}. Results are summarized in Table~\ref{tab:Fluxes_Comparison}, and compare favorably both between different implementations with the BHPT tools in vacuum.
\begin{figure}[h]
	\includegraphics[scale=0.3]{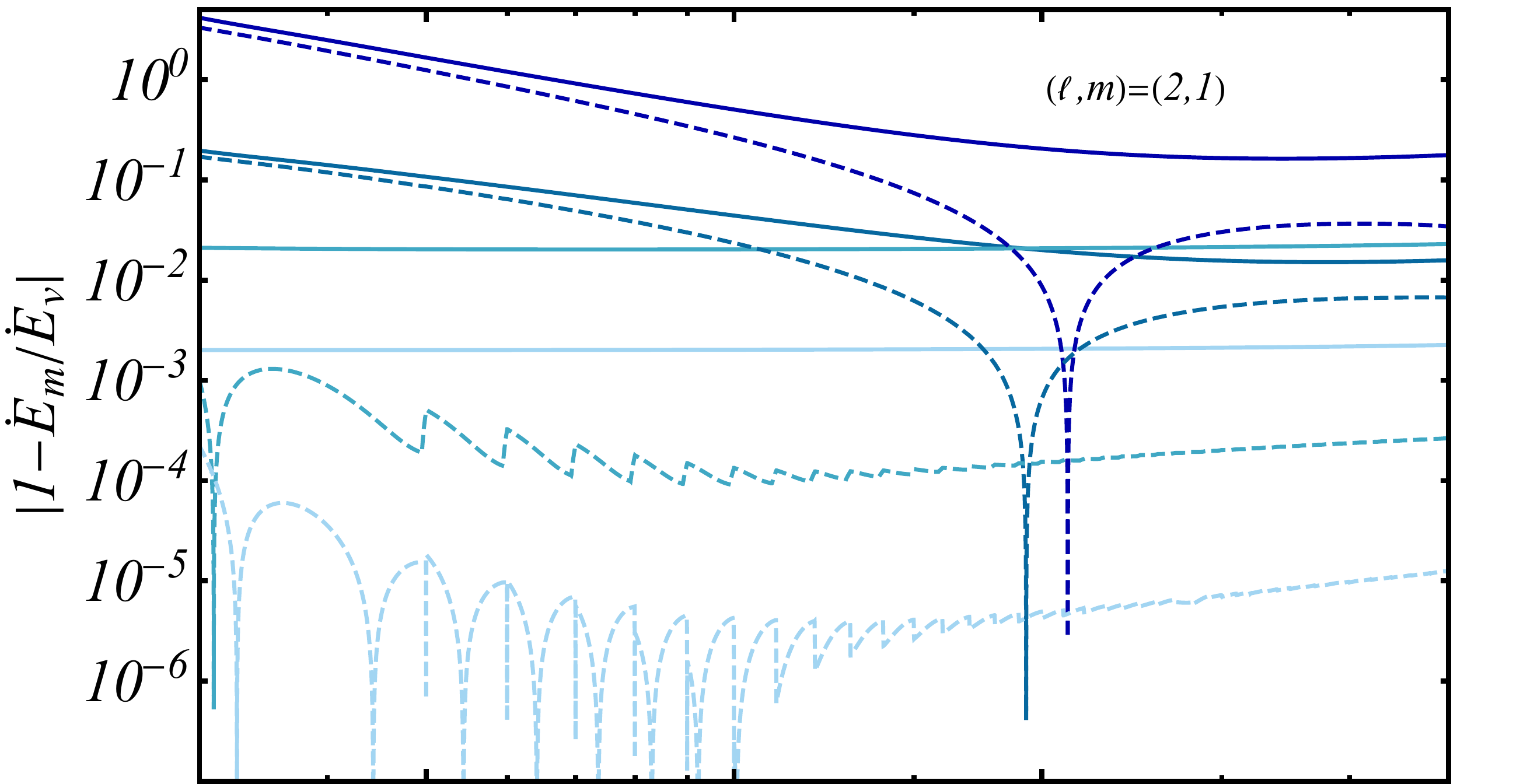}\\
	\includegraphics[scale=0.3]{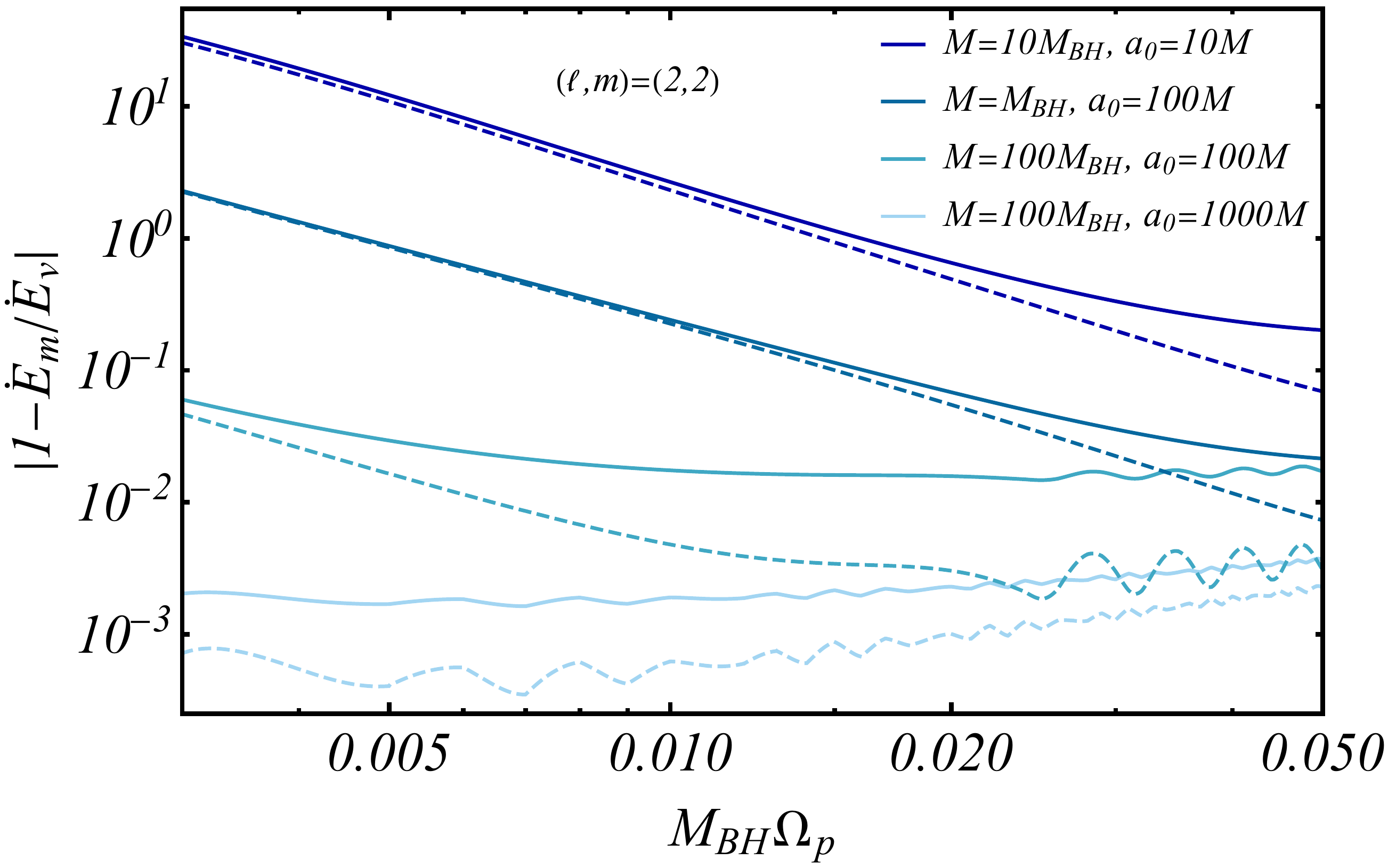}
	\caption{\textit{Top panel:} Relative difference between the energy flux of the $\ell=2$, $m=1$ mode emitted by the EMRI for different GBH configurations ($\dot{E}_m$) and in vacuum ($\dot{E}_v$), as a function of GW frequency (solid lines). Frequency range corresponds to a secondary location $r_p = 50 M_\text{BH}$ down to $r_p = 6 M_\text{BH}$. Dashed lines show the vacuum result redshifted according to Eq.~\eqref{eq:RescalingOmega}. \textit{Bottom panel:} Same as the top panel but for the $\ell=m=2$ mode.}
	\label{fig:Flux_Axial}
\end{figure}
It is clear from Table~\ref{tab:Fluxes_Comparison} that, for fixed BH mass, the fluxes are smaller in the presence of a halo. However, given that the binary sits at a nontrivial gravitational potential set by the halo, decreasing fluxes may amount to a redshift effect. We focus on realistic environments, where $M_\text{BH} \ll M \ll a_0 $. To linear order in $M/a_0$, $dr/dr_* \approx \left(1-M/a_0 \right)dr/dr_*^{\rm vac}$ where $r_*^{\rm vac}$ is the tortoise coordinate in a Schwarzschild geometry. Additionally, for compact EMRIs ($r_p \sim 10M_\text{BH}$), $S^\text{ax} \approx \left(1-3M/a_0 \right) S^\text{ax}_\text{vac}$. Combining these, expanding Eq.~\eqref{eq:master_axial} to linear order in $M/a_0$ one finds
\be
\frac{d^2 \psi^{\rm ax}}{d(r_{*}^{\rm vac})^2}  +
\left( \frac{\omega^2}{\gamma^2} -V^{\rm ax}_\text{Schw}\right)\psi^{\rm ax}= \gamma  S^\text{ax}_\text{Schw} \,,
\label{eq:MasterAxialExpanded}
\ee
where $\gamma=1-M/a_0$ is a redshift factor. Thus, to linear order in $\gamma$ the axial signal from a GBH is identical to that from a Schwarzschild BH, with redshifted frequency and mass; in other words, the two setups are equivalent with the identification
\be
\left(\Omega^{\rm vac}_p,\omega^{\rm vac},m^{\rm vac}_p\right) \rightarrow \left(
\frac{\Omega_p}{\gamma},\frac{\omega}{\gamma},\gamma m_p\right) \label{eq:RescalingOmega}
\ee
Axial perturbations do not couple to matter perturbations, and a simple propagation redshift seems adequate. The polar sector is more involved, and requires numerical study.

In Fig.~\ref{fig:Flux_Axial}, we present numerical results that confirm this picture, showing fluxes as a function of the frequency of the GWs being measured by a distant stationary observer. For axial modes ($\ell=2, m=1$), the differences between a vacuum and non-vacuum environment are seemingly large, but as can be seen in Fig.~\ref{fig:Flux_Axial}, fluxes from a GBH are indeed well described by redshifted fluxes in vacuum. The agreement is all the better for larger halo mass $M$, smaller compactness $M/a_0$. For galactic configurations, it leads to relative differences that are extremely small. 

Note that for small scales, $a_0 \omega\lesssim 1$, the radiation wavelength is larger than the halo itself, and redshift is suppressed. At large frequencies redshifted vacuum fluxes are an excellent description of our results, for axial perturbations. Indeed, we also find that quasinormal modes conform to such a description since they are high-frequency phenomena in this setup~\cite{Long_version}.

Polar fluctuations are coupled to the fluid, as we saw, and a naive redshift is not sufficient to describe GW generation and propagation. Figure~\ref{fig:Flux_Axial}
shows one of our exciting findings: polar perturbations are less prone to redshift effects, even in regions of parameter space corresponding to large, near-galactic scales. Thus, together with the axial sector they're able to break possible degeneracies, with sensitive, low-frequency detectors.
 
Independently of that, our results clearly indicate the ability of GW astronomy to strongly constrain smaller scale matter distributions around BHs. At $\omega M_{\rm BH}=0.02$, the relative flux difference between a vacuum and a GBH with $M=0.1M_{BH}$ and $a_0=10^2M,10^3M$ is $\sim 10\%,1\%$ respectively. These numbers are within reach of next generation detectors \cite{Bonga:2019ycj}. Compare with GRAVITY's constraints on the environment of the Sgr A$^*$ star \cite{GRAVITY:2021xju}, but note that GW astronomy allows similar constraints for a large number of sources.

\noindent{\bf{\em Discussion.}}
Our work serves as a proof-of-concept for the ability to study environmental effects in GW physics at a full relativistic level. A natural next step is to apply it to other environments, for example by taking input from recent GRMHD simulations of accretion~\cite{Derdzinski:2020wlw, Zwick:2021dlg}, or to add rotation to the BH.

The application of our relativistic framework to galactic EMRIs immersed in a halo shows that environments can easily de-stabilize the BH spectra, as had recently been suggested with toy models \cite{Barausse:2014tra,Jaramillo:2021tmt,Destounis:2021lum,Cheung:2021bol,Berti:2022xfj}; it is unknown at this point if environmental resonances can be excited by supermassive BHs, long-before merger; however, our results show how the coupling to the environment changes GW generation and propagation.

Nonetheless, there are important issues that remain to be answered. The energy flux emitted in GWs contains contributions directly from the binding energy of the binary but also from the environment. It is unclear if energy balance arguments alone are sufficient to evolve such systems, even in an adiabatic approach, or if self-force methods \cite{Barack:2018yvs} are necessary, and whether they too need to be modified to take environments fully into account. This aspect is of particular relevance if the binary is able to resonantly excite the proper modes of the environment. In addition to energy carried by GWs, there will also be viscous heating, which can be included in the formalism. We plan to address some of these problems in future work.  

\begin{acknowledgments}
We thank all the participants of the ``EuCAPT Workshop: Gravitational wave probes of black hole environments'' in Rome, David Hilditch, and Rodrigo Vicente for useful and lively discussions. 
V.C.\ is a Villum Investigator and a DNRF Chair, supported by VILLUM FONDEN (grant no.~37766) and by the Danish Research Foundation. V.C.\ acknowledges financial support provided under the European
Union's H2020 ERC Advanced Grant ``Black holes: gravitational engines of discovery'' grant agreement
no.\ Gravitas–101052587. F. D. acknowledges financial support provided by FCT/Portugal through grant No. SFRH/BD/143657/2019. R.P.M acknowledges financial support provided from STFC via grant number ST/V000551/1.
K.D. acknowledges financial support provided under the European Union's H2020 ERC, Starting Grant agreement no.~DarkGRA--757480 and the MIUR PRIN and FARE programmes (GW-NEXT, CUP: B84I20000100001).
This project has received funding from the European Union's Horizon 2020 research and innovation programme under the Marie Sklodowska-Curie grant agreement No 101007855.
We thank FCT for financial support through Projects~No.~UIDB/00099/2020 and UIDB/04459/2020.
We acknowledge financial support provided by FCT/Portugal through grants 
2022.01324.PTDC, PTDC/FIS-AST/7002/2020, UIDB/00099/2020 and UIDB/04459/2020.
We acknowledge financial support provided by FCT/Portugal through grants PTDC/MAT-APL/30043/2017 and PTDC/FIS-AST/7002/2020.
\end{acknowledgments}

\newpage
\begin{widetext}


\begin{center}
{\Large \bf Supplemental material}
\end{center}

\section{Perturbation equations}
We will impose the Regge-Wheeler gauge~\cite{Regge:1957td,Zerilli:1970se}, such that the metric components obey
\begin{align}
g_{\theta\phi}=0\quad \ ,&\quad g_{\phi\phi}=g_{\theta\theta}\sin^2\theta\ ,\nonumber\\
\partial_{\phi}(g_{t\phi}/\sin\theta)&+\partial_\theta(g_{t\theta}\sin\theta)=0\ ,\nonumber\\
\partial_{\phi}(g_{r\phi}/\sin\theta)&+\partial_\theta(g_{r\theta}\sin\theta)=0\ .
\end{align}
With this choice the non-vanishing metric perturbations $g^{(1)}_{\mu\nu}=g^{(1)\tn{axial}}_{\mu\nu}+g^{(1)\tn{polar}}_{\mu\nu}$ of Eq.~(3) in the main text take the following form
\begin{equation}
g^{(1)\tn{axial}}_{\mu\nu}(t,r,\theta,\phi)=\sum_{\ell=2}^{\infty}\sum_{m=-\ell}^\ell\frac{\sqrt{2\ell(\ell+1)}}{r} [ih_{1}^{\ell m }(t,r)c_{\ell m,\mu\nu}(r,\theta,\phi) \nonumber 
-h_{0}^{\ell m}(t,r)c^{0}_{\ell m,\mu\nu}(r.\theta,\phi)]\ ,
\end{equation}
\begin{align}
g^{(1)\tn{polar}}_{\mu\nu}(t,r,\theta,\phi)=\sum_{\ell=2}^{\infty}\sum_{m=-\ell}^\ell\bigg[a&H^{\ell m }_0(t,r)a^{0}_{\ell m,\mu\nu}(\theta,\phi)-i\sqrt{2}H^{\ell m}_1(t,r)a^{1}_{\ell m,\mu\nu}(\theta,\phi)
\nonumber\\
&+\frac{H_2^{\ell m }(t,r)}{b}a_{\ell m,\mu\nu}(\theta,\phi)
+\sqrt{2}r^2K^{\ell m }(t,r)g_{\ell m,\mu\nu}(r,\theta,\phi)\bigg]\ .\label{expansion}
\end{align}
\end{widetext}
with $\{c_{\ell m,\mu\nu},c^{0}_{\ell m,\mu\nu},a^{0}_{\ell m,\mu\nu}, a^{0}_{\ell m,\mu\nu},a_{\ell m,\mu\nu},g_{\ell m,\mu\nu}\}$ being six out of ten tensor spherical harmonics, which work as a basis to expand any tensor field~\cite{Sago:2002fe}.

As also mentioned in the main text, the perturbations of the stress-energy tensor associated with the environment, $T^\tn{env(1)}_{\mu\nu}$,  can be treated in the same way as for the metric. $T^\tn{env(1)}_{\mu\nu}$ encodes the fluid perturbations through the energy-density changes (Eq.~(4) in the main text):
\begin{align}
p^{(1)}_r(t,r,\theta,\phi)&=\sum_{\ell=2}^\infty\sum_{m=-\ell}^\ell\delta p_{r,\ell m}(t,r)Y_{\ell m}(\theta,\phi)\label{math:preperttang}\ ,\\
p^{(1)}_t(t,r,\theta,\phi)&=\sum_{\ell=2}^\infty\sum_{m=-\ell}^\ell\delta p_{t,\ell m}(t,r)Y_{\ell m}(\theta,\phi)\label{math:prepertrad}\ ,\\
\rho^{(1)}(t,r,\theta,\phi)&=\sum_{\ell=2}^\infty\sum_{m=-\ell}^\ell\delta \rho_{\ell m}(t,r)Y_{\ell m}(\theta,\phi)\label{math:rhopert}\ ,
\end{align}
In the addition to energy and density, we also need to perturb the fluid's 4-velocity $u^{\mu}_{(1)}$. These are described by three functions $\{U_{\ell m}(t,r),V_{\ell m}(t,r),W_{\ell m}(t,r)\}$ and the condition $u^2=-1$ (up to first order), such that:
\begin{align}
u^{t}_{(1)}&=\frac{1}{2a^{1/2}}\sum_{\ell=2}^\infty\sum_{m=-\ell}^\ell H^{\ell m}_0 Y_{\ell m}\,,\\
u^{r}_{(1)}&=\frac{a^{1/2}}{b}\frac{1}{4\pi\kappa}\sum_{\ell=2}^\infty\sum_{m=-\ell}^\ell W_{\ell m}Y_{\ell m}\,,\\
u^{\theta}_{(1)}&=\frac{a^{1/2}}{4\pi\kappa r^2}\sum_{\ell=2}^\infty\sum_{m=-\ell}^\ell\bigg[V_{\ell m}\partial_\theta -\frac{U_{\ell m}}{\sin\theta}\partial_\phi\bigg]Y_{\ell m}\ ,\\
u^{\phi}_{(1)}&=\frac{a^{1/2}}{4\pi\kappa r^2\sin^2\theta}\sum_{\ell=2}^\infty\sum_{m=-\ell}^\ell\bigg[V_{\ell m}\partial_\phi +\frac{U_{\ell m}}{\sin\theta}\partial_\theta\bigg]Y_{\ell m}\ ,
\end{align}
where we suppressed the coordinate dependence to improve readability. $\kappa(r)$ is a generic function usually taken to be $\kappa(r)=\rho^{(0)}(r)+p^{(0)}(r)$. Since we are interested in studying an anisotropic background with vanishing radial pressure we take without loss of generality $\kappa(r)=\rho^{(0)}(r)+p^{(0)}_t(r)$ when needed. As mentioned in the main text, assuming a barotropic equation of state provides a further relation between the density and pressure perturbations:
\beq
\delta p_{r,\ell m}&=&c_{s_r}^2\delta \rho_{\ell m}\ , \\
\delta p_{t,\ell m}&=&c_{s_t}^2\delta \rho_{\ell m}\ ,
\eeq
where in general $c_{s_r}=c_{s_r}(r)$ and $c_{s_t}=c_{s_t}(r)$ are functions of $r$.

The components of $T_{\mu\nu}^{\tn{env}(1)}$ as a function of the fluid perturbations are then
\begin{align}
T_{tt}^{\tn{env}(1)}=&a\sum_{\ell=2}^\infty\sum_{m=-\ell}^\ell(\delta\rho_{\ell m}-H^0_{\ell m}\rho^{(0)})Y_{\ell m}\ ,\nonumber\\
T_{tr}^{\tn{env}(1)}=&-\sum_{\ell=2}^\infty\sum_{m=-\ell}^\ell\left[\frac{a\left(\rho^{(0)} + p^{(0)}_r\right)}{4\pi\kappa  b^2}W_{\ell m}+H^{1}_{\ell m} \rho^{(0)}\right] Y_{\ell m}\ ,\nonumber\\
T_{t\theta}^{\tn{env}(1)}=&
\frac{a\left(\rho^{(0)} + p^{(0)}_t\right)}{4\pi\kappa}\sum_{\ell=2}^\infty\sum_{m=-\ell}^\ell\left[\csc\theta  U_{\ell m} \partial_{\phi}-V_{\ell m}\partial_{\theta}\right]Y_{\ell m}\ ,\nonumber\\
T_{t\phi}^{\tn{env}(1)}=&
-\frac{a\left(\rho^{(0)} + p^{(0)}_t\right)}{4\pi \kappa}\sum_{\ell=2}^\infty\sum_{m=-\ell}^\ell\left[V_{\ell m} \partial_{\phi}+U_{\ell m}\sin\theta\partial_{\theta}\right]Y_{\ell m}\ ,\nonumber\\
T_{rr}^{\tn{env}(1)}=& \frac{1}{b}\sum_{\ell=2}^\infty\sum_{m=-\ell}^\ell \left( p^{{(0)}}_r H^2_{\ell m}+
\delta p_{r,\ell m} \right)Y_{\ell m}\nonumber\ ,\\ 
T_{\theta\theta}^{\tn{env}(1)}=&r^2 \sum_{\ell=2}^\infty\sum_{m=-\ell}^\ell(p^{{(0)}}_t K_{\ell m}+
\delta p_{t,\ell m})Y_{\ell m}\nonumber\ ,\\ 
T_{\phi\phi}^{\tn{env}(1)}=&T_{\theta\theta}^{\tn{halo}(1)}\sin^2\theta\ .
\end{align}

The stress-energy tensor of the EMRI's secondary, $T^{\mu\nu}_p$ (Eq.~(6) in the main text), which sources the metric and fluid perturbations can also be decomposed in terms of the tensor harmonics basis previously introduced \cite{Sago:2002fe}: 
\begin{widetext}
\begin{align}
T^{p}_{\mu\nu}(t, r, \theta, \phi)=\sum_{\ell=2}^{\infty}\sum_{m =-\ell}^{\ell}
\bigg[&{\cal A}^{0}_{\ell m }a^{0}_{\ell m,\mu\nu}(\theta,\phi)+{\cal A}^{1}_{\ell m}a^{1}_{\ell m,\mu\nu}(\theta,\phi)
+{\cal A}_{\ell m }a_{\ell m,\mu\nu}(\theta,\phi)+{\cal B}^{0}_{\ell m }b^{0}_{\ell m,\mu\nu}(r,\theta,\phi)\nonumber\\
&+{\cal B}_{\ell m }b_{\ell m,\mu\nu}(r,\theta,\phi)+{\cal Q}^{0}_{\ell m }c^{0}_{\ell m,\mu\nu}(r,\theta,\phi)+{\cal Q}_{\ell m }c_{\ell m,\mu\nu}(r,\theta,\phi)\nonumber\\
&+{\cal D}_{\ell m }d_{\ell m,\mu\nu}(r,\theta,\phi)+{\cal G}_{\ell m}g_{\ell m,\mu\nu}(r,\theta,\phi)+{\cal F}_{\ell m }f_{\ell m,\mu\nu}(r,\theta,\phi)\bigg]\ .\label{harmonicexp}
\end{align}
\end{widetext}

The expansion coefficients 
$({\cal A}^{0}_{\ell m },{\cal A}^{1}_{\ell m },\ldots,{\cal F}_{\ell m })$ 
can be computed by projecting each tensor harmonics on the stress-energy tensor, e.g. $Q^{0}_{\ell m }=(c^{0}_{\ell m,\mu\nu},T_{\mu\nu})$, where the operator $(\ ,\ )$ acts on two generic basis components $A,B$ such that:
\begin{equation}
(A,B)=\int\int \eta^{\mu\rho}\eta^{\nu\sigma}A^{\star}_{\mu\nu}B_{\rho\sigma}d\Omega\ ,
\end{equation}
where $\eta^{\mu\nu}$ is the Minkowski metric in spherical 
coordinates and the superscript $\star$ denotes complex conjugation.

\section{The secondary stress-energy tensor for circular orbits}\label{Appendix:tensorcoefficients}

We now focus on circular orbits at radius $r_p$ which due to spherical symmetry can be taken to be at the equatorial plane ($\theta_p=\pi/2$) without loss of generality. The 4-velocity of the secondary is
\beq
u^\mu_p = \left( \frac{E_p}{a_p} , 0, 0 , \frac{L_p}{r_p^2} \right)\, , 
\eeq
where $a_p=a(r_p)$, and $E_p$ and $L_p$ are, respectively, the energy and angular momentum per unit rest mass of the small body and determined by
\beq E_p = \dfrac{a_p}{\sqrt{a_p - r_p^2 \Omega_p^2}},  \quad 
L_p = \dfrac{\Omega_p r_p^2}{\sqrt{a_p - r_p^2 \Omega_p^2}}.
\eeq
with the angular orbital frequency $\varphi_p(t) = \Omega_p \, t$ given by
\beq
\Omega_p = \sqrt{\dfrac{a'	_p}{2r_p}}.
\eeq

For this orbital configuration, the tensor harmonics expansion~\eqref{harmonicexp} for the secondary stress-energy tensor greatly simplifies. We have ${\cal A}_{\ell m}={\cal A}_{\ell m}^{(1)}={\cal B}_{\ell m}={\cal Q}_{\ell m}=0$, while the non vanishing coefficients can be expressed in terms of the orbital parameters as follows:
\beq
{\cal A}_{\ell m }^{0}&=&\frac{m_p\sqrt{a b}E_p}{r^2} Y^\star_{\ell m }\,\delta_{r}\ ,\nonumber\\
{\cal B}_{\ell m }^{0}&=&\frac{m_pi\sqrt{ab}L_p}{r^3\sqrt{(n+1)}}\,\delta_{r} \,\partial_\phi Y^\star_{\ell m }\ ,\nonumber\\
{\cal Q}_{\ell m }^{(0)}&=&-\frac{m_p\sqrt{ab}L_p}{r^3\sqrt{(n+1)}}\,\delta_{r} \,\partial_\theta Y^\star_{\ell m }\ ,\nonumber\\
{\cal G}_{\ell m }&=&\frac{m_pL_p^2\sqrt{ab}}{r^4\sqrt{2}E_p}\,\delta_{r} \, Y^\star_{\ell m}\ ,\nonumber\\
{\cal D}_{\ell m }&=&\frac{m_piL_p^2\sqrt{ab}}{E_pr^4\sqrt{2n(n+1)}}\,\delta_r \, \partial_{\theta\phi}Y^\star_{\ell m}
\ ,\nonumber\\
{\cal F}_{\ell m }&=&\frac{m_pL_p^2\sqrt{ab}}{r^42E_p\sqrt{2n(n+1)}}\, \delta_r \, (\partial_{\phi\phi}-\partial_{\theta\theta}) Y^\star_{\ell m}\ , \nonumber \\
\eeq
where 
$n=\ell(\ell+1)/2-1$, $Y^\star_{\ell m}=Y^\star_{\ell m}(\theta_p,\varphi_p)$
and $\delta_{r}=\delta(r-r_p)$.

As a final comment, we can also study perturbations in the frequency domain by applying a Fourier transform on both sides of the field equations, such that the generic frequency dependent perturbation $\psi(\omega,r)$, is given by
\be
\psi^{\ell m}(\omega,r)=\frac{1}{\sqrt{2\pi}}\int_{-\infty}^{\infty}\,dt\,e^{i\omega t} \psi^{\ell m}(t, r)\ .
\ee
%

\section{Master equations}\label{sec:masterequations}

Having defined the decomposition of metric and stress-energy tensor perturbations in terms of radial and angular variables, we can now derive a set of partial differential equations for the axial and polar  sectors. For the sake of clarity, hereafter, we will drop the sum over the multipolar indices. Also to facilitate reproducibility, we will particularize for the case of vanishing background radial pressure, $p^{(0)}_r=0$, but the same steps can be followed for the more general case. We will often make use of the following $0-th$ order relations between the background quantities to simplify the expressions
\beq
a' &=& \frac{2 a}{r^2} \frac{m}{1-2m/r} \, , \\
m' &=& 4\pi r^2 \rho \, , \\
p_t^{(0)} &=& \frac{m}{2(r-2 m)} \rho^{(0)} \, .
\eeq
%

\subsection{Axial sector}\label{sec:axial}

The problem for axial perturbations requires to find a solution for 
$h^{\ell m}_0$, $h_1^{\ell m}$ and $U^{\ell m}$. We start by rewriting $\partial h^{\ell m}_0/\partial t$ using the combination of 
${\cal E}_{\theta\theta}-{\cal E}_{\phi\phi}/\sin^2\theta$, where 
\begin{equation}
{\cal E}_{\mu\nu}=G^{(1)}_{\mu\nu}-8\pi(T^{\rm env(1)}_{\mu\nu}+T^p_{\mu\nu})\ ,
\end{equation}
and $G^{(1)}_{\mu\nu}$ is the perturbed Einstein tensor
\begin{align}
	\frac{\partial h^{\ell m}_0}{\partial t}=a\, b \frac{dh^{\ell m}_1}{dr}
	&+\frac{a(1-b+r b')}{2r}h^{\ell m}_{1}\nonumber\\-
	&i \frac{4\sqrt{2}\pi r^2a}{\sqrt{n(1+n)}}{\cal D}^{\ell m}\ .
\end{align}
By defining the new variable $\chi^{\ell m}=(a b)^{1/2}/r h_1^{\ell m}$, 
the ${\cal E}_{r\theta}$ component provides a second order non-homogeneous 
differential equation for $\chi^{\ell m}$. In terms of the generalized 
tortoise coordinate $dr_*/dr=(a\, b)^{1/2}$, the master equation for the axial metric perturbations can be written as 
\begin{equation}
\left[ - \frac{\partial^2 }{\partial t^2} + \dfrac{\partial^2 }{\partial r_*^2} - V^{\rm ax}\right]\chi=  S^{\rm ax}\ ,\label{eq:master_axial} 
\end{equation}
where the potential reads 
\begin{equation}
V^{\rm ax} = \dfrac{a}{r^2}\left[\ell(\ell+1) - \dfrac{6m(r)}{r} + m'(r)  \right]\ ,
\end{equation}
and the source term is
\beq
\label{eq:Source}
S^{\rm ax} = G\delta(r- r_p) + F \delta'(r-rp) \, ,
\eeq
with
\beq
G &= &\chi_p  \, \dfrac{a}{2r^4} \bigg( 3\,r\, a'\, b + r\, a\, b' - 8 \,a\,b \bigg) \, , \\
F &=& \chi_p  \dfrac{a^2\, b}{r^3} \, , \\
\chi_p &=&  m_p \frac{4\pi}{n(n+1)} \, \dfrac{L_p^2}{E_p} e^{-i m \Omega_p t} X_{lm}^*\left(\frac{\pi}{2},0\right) \, , 
\eeq
and
\beq
X_{lm}^*\left(\frac{\pi}{2},0\right) &=& - im \frac{\partial}{\partial\theta}Y_{\ell m}^*\Big|_{\theta=\frac{\pi}{2},\varphi=0} \, .
\eeq
This last $m$ is the azimuthal number (we always write the mass function as $m(r)$). Note that by the symmetry of the spherical harmonics, only modes with $l+m$ odd have non-zero contributions to the axial sector (and only modes with $l+m$ even have non-zero contributions to the polar sector) 

Finally, the fluid perturbation $U^{\ell m}$ can be determined in terms of 
$\chi^{\ell m}$ and its first derivative using the component ${\cal E}_{t\theta}$ of the field's equations. 

\subsection{Polar sector}\label{sec:polar}

For the polar sector we present two equivalent formulations. In the time-domain, we derive a system of 3 ``wavelike'' equations with an additional constraint equation, while in the frequency-domain we obtain 5 first-order  ODEs.

\subsubsection{Time-domain: 3 second-order ``wavelike'' PDEs}

First, let us rewrite 
\beq
H_0(t,r) &=& K(t,r) +  \frac{r}{a} S(t,r) \, , \\ 
H_1(t,r) &=&  \frac{r}{a} \tilde{H}_1(t,r) \, .
\eeq

\begin{itemize}

\item $\mathcal{E}_{\theta\theta} - \mathcal{E}_{\varphi\varphi}/\sin^2 \theta$ gives an algebraic relation for $H_2$
\beq
H_2 &=& K +  \frac{r}{a} S - \frac{16 \pi r^2}{\sqrt{2n(n+1)}} \,  \, \mathcal{F}_{\ell m} \, , 
\eeq

\item $\mathcal{E}_{r\theta}$ gives a dynamical equation for $\tilde{H}_1$ which we use to substitute $\partial \tilde{H}_1/\partial t$ and $\partial^2 \tilde{H}_1/\partial t\partial r_*$ when necessary 

\beq
\frac{\partial \tilde{H}_1}{\partial t} &=& \sqrt{\frac{a}{1-2m/r} } \frac{\partial S}{\partial r_*} + \frac{a}{r}S + \frac{2 \, a^2\, m}{r^3(1-2m/r)}K  \nonumber \\
&-& \frac{16 \pi}{\sqrt{2n(n+1)}} a^2 \frac{r-m}{r-2m} \mathcal{F}_{\ell m} \, , 
\eeq

\begin{widetext}
\item $\mathcal{E}_{tt} = 8 \pi T_{tt}$ yields a constraint between $K$, $S$ and $\delta \rho$, 
\beq
\frac{\partial^2 K}{\partial r_*^2} &=&  \sqrt{\frac{1-2m/r}{a}} \frac{\partial S}{\partial r_*} + \sqrt{\frac{a}{1-2m/r}} \left( \frac{5m}{r^2} - \frac{2}{r}  \right)\frac{\partial K}{\partial r_*} \nonumber \\
&+& a \left(\frac{\ell(\ell +1)}{r^2} - 8 \pi \rho \right)K + \left( \frac{\ell(\ell+1)+4}{2r} -\frac{4m}{r^2} -8\pi r \rho\right)S   - 8 \pi a \,\delta \rho \nonumber \\
&-& 8 \pi A^0_{\ell m} + \frac{8\pi\, a}{\sqrt{2n(n+1)}}  \left( 16 \pi r^2 \rho + \frac{8m}{r} - \ell \left(\ell +1 \right) - 6 \right)  \mathcal{F}_{\ell m}  - \frac{16\pi \, a r}{\sqrt{2n(n+1)}}  \left(1 - \frac{2m}{r} \right) \frac{\partial \mathcal{F}_{\ell m}}{\partial r} \, , \nonumber \\
\label{eq:HamiltonianConstraint}
\eeq
\end{widetext}
\begin{widetext}
\item $\mathcal{E}_{tt} - a\, b\, \mathcal{E}_{rr}$ gives the first second-order ``wavelike'' equation for $K$

\beq
\label{eq:K}
&-&\frac{\partial^2 K}{\partial^2 t} + \frac{\partial^2 K}{\partial^2 r_*} + \frac{2}{r} \sqrt{a(1-2m/r)} \frac{\partial K}{\partial r_*} + \frac{a}{r^2}\left(8\pi r^2 \, \rho + \frac{4m}{r} - \ell(\ell+1) \right) K \nonumber \\
&=& -8 \pi a \left(1 - c_{s_r}^2 \right) \, \delta \rho + \frac{2}{r}\left(1 - \frac{2m}{r} -  4 \pi r^2 \rho \right)S \nonumber \\
&-& \frac{8 \pi  \, a}{{\sqrt{2n(n+1)}}} \left( 4 + \ell \left(\ell +1 \right) - \frac{4m}{r} - 16 \pi r^2 \rho \right) \mathcal{F}_{\ell m} - \frac{16\pi  \, a\, r}{\sqrt{2n(n+1)}}\left(1 - \frac{2m}{r} \right) \frac{\partial \mathcal{F}_{\ell m}}{\partial r}  - 8 \pi \mathcal{A}^0_{\ell m} \, .
\label{eq:WaveEquationK}
\eeq
\end{widetext}	
\begin{widetext}
\item $\mathcal{E}_{\theta\theta} + \mathcal{E}_{\varphi\varphi}/\sin^2 \theta$  gives another second-order ``wavelike'' equation for $S$, where one needs to use the previous equations to substitute the necessary derivatives
\beq
&-&\frac{\partial^2 S}{\partial^2 t} + \frac{\partial^2 S}{\partial^2 r_*} + \frac{a}{r^2}\left(4\pi r^2 \, \rho + \frac{2m}{r} - l(l+1) \right) S \nonumber \\
&=& \frac{4a^2}{r^4 \left(r - 2 m \right)} \left( 3\, m \left( r + 2\pi r^3 \rho \right) - 7 m^2 - 4\pi r^4 \rho \right) K - 16\pi \frac{a^2}{r}\left(c_{s_r}^2 - c_{s_t}^2  \right) \delta \rho \nonumber \\
&-& \frac{8 \pi}{\sqrt{2n(n+1)}} \frac{a^2}{r^3 \left(1-2m/r\right)}\left(20 m^2 - 2 \left(\ell^2 + \ell + 6 \right)m \, r + \ell \left(\ell + 1\right)r^2  \right) \mathcal{F}_{\ell m} \nonumber \\
&-& \frac{16 \pi \, a}{\sqrt{2n(n+1)}} \left ( \frac{a}{r}\left(m - r \right) \frac{\partial \mathcal{F}_{\ell m}}{\partial r}   + r \frac{\partial^2 \mathcal{F}_{\ell m}}{\partial t^2}\right ) + 8\pi \sqrt{2} \, \frac{a^2}{r} \mathcal{G}_{\ell m} \, .
\label{eq:WaveEquationS}
\eeq

The final ``wavelike'' equation for $\delta \rho$ is obtained from the conservation of the perturbed stress-energy tensor of the surrounding fluid. 
\item $\nabla_\mu T^{\mu \theta} = 0$

\beq
\frac{\partial V}{\partial t} &=& 2\pi \rho\left(K + \frac{r}{a}S \right) - 4 \pi c_{s_t}^2 \delta \rho + \frac{32 \pi^2 \, r^2 \, p_t^{0} }{{\sqrt{2n(n+1)}}} \, \mathcal{F}_{\ell m} \, . 
\eeq
\item $\nabla_\mu T^{\mu r} = 0$ 
\beq
&&\frac{\partial W}{\partial t} = - 2\pi \left(2r -3m\right) \left(1-\frac{2m}{r}\right)\frac{\rho}{a^2}\frac{\partial \tilde{H}_1}{\partial t}	- \frac{2\pi}{r} \left(2r - 3m \right)\sqrt{\frac{1-2m/r}{a}}  c_{s_r}^2 \frac{\partial \delta \rho}{\partial r_*} \nonumber \\
&+& \pi \left(2r -3m\right) \frac{\rho}{a} \left( \sqrt{\frac{1-2m/r}{a}} \frac{\partial S }{\partial r_*} + \frac{r- m}{r^2} \sqrt{\frac{a}{1-2m/r}}\frac{\partial K }{\partial r_*} \right) \nonumber \\
&+& \frac{\pi}{r^2} \left(2r - 3m\right)\left(\left(r - 4m\right)\frac{\rho}{a}S - 2\left( 2\left(c_{s_r}^2 - c_{s_t}^2\right) + (1 - 3\, c_{s_r}^2 + 4 \, c_{s_t}^2)\frac{m}{r}  \right) \delta \rho  \right) \, .
\eeq
\end{widetext}
\begin{widetext}
\item $\partial_t \left( \nabla_\mu T^{\mu t} \right) = 0$ and using all the previous equations we finally arrive at
\beq
&-&\frac{\partial^2 \delta \rho}{ \partial t^2} + c_{s_r}^2 \frac{\partial^2 \delta \rho}{\partial r_*^2} + \sqrt{\frac{a}{1-2m/r} } \left( \frac{2}{r}\left(2 c_{s_r}^2 - c_{s_t}^2 \right) + \left( 1 - 5 c_{s_r}^2 + 4 c_{s_t}^2  \right) \frac{m}{r^2} + 2 \left(1-\frac{2m}{r}\right)c_{s_r} c'_{s_r}  \right)\frac{\partial \delta \rho}{\partial r_*} \nonumber \\
&+& \frac{a}{1-2m/r}\Bigg( \frac{2 c_{s_r}^2 - c_{s_t}^2 \left(\ell^2 + \ell + 2 \right)}{r^2} + \frac{2m}{r^3}\left( c_{s_t}^2 \ell \left(\ell + 1\right) + \left(1 - 3 c_{s_r}^2 + 4 c_{s_t}^2 \right) \frac{m}{r} \right) \nonumber \\
&+& 8\pi \rho \left( 1 + 2 c_{s_t}^2 + \left(c_{s_r}^2 - 4 c_{s_t}^2 - 1 \right) \frac{m}{r} \right) + \frac{2 }{r^2}\left(1-\frac{2m}{r}\right)\left(2r - 3m\right)c_{s_r} c'_{s_r} - \frac{4 }{r}\left(1 -\frac{2m}{r} \right)^2 c_{s_t} c'_{s_t}  \Bigg) \delta \rho \nonumber \\
&=& - \frac{r}{2}\frac{\partial \rho}{\partial r} \sqrt{\frac{1-2m/r}{a}}  \frac{\partial S}{\partial r_*} - \frac{1}{2r}\sqrt{\frac{a}{1-2m/r}}\left( (m-r) \frac{\partial \rho}{\partial r} - \left( \frac{m}{r} + 4 \pi r^2 \rho \right) \frac{\rho}{1-2m/r}  \right)\frac{\partial K}{\partial r_*} \nonumber \\
&-& \frac{1}{2}\left( \left( 8\pi r \rho - \frac{m}{r^2}\ell \left(\ell + 1\right) \right)\frac{\rho}{1-2m/r}  + \frac{\partial \rho}{\partial r} \right) S - \frac{a}{2r^2} \left( \left( 8 \pi r^2 \rho - \ell \left(\ell +1 \right)\frac{m}{r} \right) \frac{\rho}{1-2m/r} + 4 m \frac{\partial \rho}{\partial r}  \right) K\nonumber \\
&-& 4 \pi\sqrt{2}a\rho \, \mathcal{G}_{\ell m} - 4\pi \rho \frac{1-m/r}{1-2m/r} \, A^0_{\ell m} - \frac{8 \pi \, a}{\sqrt{2 n(n+1)}}  \left( \left( \ell (\ell + 1) \frac{m}{r} - 8 \pi r^2 \rho \right)\frac{\rho}{1-2m/r} + 2(m - r)\frac{\partial \rho}{\partial r} \right) \mathcal{F}_{\ell m} \, . \nonumber \\
\label{eq:WaveEquationRho}
\eeq
\end{widetext}
\end{itemize}

In the matrix formulation presented in the main text
\be
\hat {\cal L} \vec{\phi}= \hat {\textbf{B}} \vec{\phi}_{,r^*} + \hat{\textbf{A}}\vec{\phi}+\vec{S}_1\,,
\ee
where $\vec{\phi}=(S,K,\delta\rho)$, the non-zero matrix coefficients are
\begin{widetext}
\beq
B_{22} &=& - \frac{2}{r} \sqrt{a(1-2m/r)} \, , \,  B_{33} =  -\sqrt{\frac{a}{1-2m/r} } \left( \frac{2}{r}\left(2 c_{s_r}^2 - c_{s_t}^2 \right) + \left( 1 - 5 c_{s_r}^2 + 4 c_{s_t}^2  \right) \frac{m}{r^2} + 2 \left(1-\frac{2m}{r}\right)c_{s_r} c'_{s_r} \right) \, , \nonumber \\
B_{31} &=& - \frac{r}{2}\frac{\partial \rho}{\partial r} \sqrt{\frac{1-2m/r}{a}}  \, , \, B_{32} = - \frac{1}{2r}\sqrt{\frac{a}{1-2m/r}}\left( (m-r) \frac{\partial \rho}{\partial r} - \left( \frac{m}{r} + 4 \pi r^2 \rho \right) \frac{\rho}{1-2m/r}  \right) \, , \nonumber \\
A_{11} &=& -\frac{a}{r^2}\left(4\pi r^2 \, \rho + \frac{2m}{r} - l(l+1) \right) \, , \, A_{12} = \frac{4a^2}{r^4 \left(r - 2 m \right)} \left( 3\, m \left( r + 2\pi r^3 \rho \right) - 7 m^2 - 4\pi r^4 \rho \right) \, , \nonumber \\
A_{13} &=& - 16\pi \frac{a^2}{r}\left(c_{s_r}^2 - c_{s_t}^2  \right) \, , \, A_{21} =  \frac{2}{r}\left(1 - \frac{2m}{r} -  4 \pi r^2 \rho \right)  \, , \, A_{22} = -\frac{a}{r^2}\left(8\pi r^2 \, \rho + \frac{4m}{r} - \ell(\ell+1) \right)\, , \nonumber \\ 
A_{23} &=& -8 \pi a \left(1 - c_{s_r}^2 \right) \, , \, A_{31} = - \frac{1}{2}\left( \left( 8\pi r \rho - \frac{m}{r^2}\ell \left(\ell + 1\right) \right)\frac{\rho}{1-2m/r}  + \frac{\partial \rho}{\partial r} \right) \, , \nonumber \\
A_{32} &=& - \frac{a}{2r^2} \left( \left( 8 \pi r^2 \rho - \ell \left(\ell +1 \right)\frac{m}{r} \right) \frac{\rho}{1-2m/r} + 4 m \frac{\partial \rho}{\partial r}  \right) \, , \nonumber \\
A_{33} &=& - \frac{a}{1-2m/r}\Bigg( \frac{2 c_{s_r}^2 - c_{s_t}^2 \left(\ell^2 + \ell + 2 \right)}{r^2} + \frac{2m}{r^3}\left( c_{s_t}^2 \ell \left(\ell + 1\right) + \left(1 - 3 c_{s_r}^2 + 4 c_{s_t}^2 \right) \frac{m}{r} \right) \nonumber \\
&+& 8\pi \rho \left( 1 + 2 c_{s_t}^2 + \left(c_{s_r}^2 - 4 c_{s_t}^2 - 1 \right) \frac{m}{r} \right) + \frac{2 }{r^2}\left(1-\frac{2m}{r}\right)\left(2r - 3m\right)c_{s_r} c'_{s_r} - \frac{4 }{r}\left(1 -\frac{2m}{r} \right)^2 c_{s_t} c'_{s_t}  \Bigg)\, , \nonumber \\
S_{1} &=& - \frac{8 \pi}{\sqrt{2n(n+1)}} \frac{a^2}{r^3 \left(1-2m/r\right)}\left(20 m^2 - 2 \left(\ell^2 + \ell + 6 \right)m \, r + \ell \left(\ell + 1\right)r^2  \right) \mathcal{F}_{\ell m} \nonumber \\
&-& \frac{16 \pi \, a}{\sqrt{2n(n+1)}} \left ( \frac{a}{r}\left(m - r \right) \frac{\partial \mathcal{F}_{\ell m}}{\partial r}   + r \frac{\partial^2 \mathcal{F}_{\ell m}}{\partial t^2}\right ) + 8\pi \sqrt{2} \, \frac{a^2}{r} \mathcal{G}_{\ell m} \, , \nonumber  \\
S_{2} &=& - \frac{8 \pi  \, a}{{\sqrt{2n(n+1)}}} \left( 4 + \ell \left(\ell +1 \right) - \frac{4m}{r} - 16 \pi r^2 \rho \right) \mathcal{F}_{\ell m} - \frac{16\pi  \, a\, r}{\sqrt{2n(n+1)}}\left(1 - \frac{2m}{r} \right) \frac{\partial \mathcal{F}_{\ell m}}{\partial r}  - 8 \pi \mathcal{A}^0_{\ell m} \, , \nonumber \\
S_{3} &=& - 4 \pi\sqrt{2}a\rho \, \mathcal{G}_{\ell m} - 4\pi \rho \frac{1-m/r}{1-2m/r} \, A^0_{\ell m} - \frac{8 \pi \, a}{\sqrt{2 n(n+1)}}  \left( \left( \ell (\ell + 1) \frac{m}{r} - 8 \pi r^2 \rho \right)\frac{\rho}{1-2m/r} + 2(m - r)\frac{\partial \rho}{\partial r} \right) \mathcal{F}_{\ell m}  \, . \nonumber \\
\eeq
\end{widetext}
%

\subsubsection{Frequency-domain: 5 first-order ODEs}

A set of first order differential equations for the metric  polar perturbations can be derived from the $t$-$r$, $t$-$\theta$ and $r$-$\theta$ components of ${\cal E}_{\mu\nu}$, which provide three inhomogeneous equations for $dH_1/dr,dK/dr$ and 
$dH_0/dr$. Moreover, the $t$ and $r$ components of the conservation equation, 
$T^{\mu\nu}{_{;\mu}}=0$ yield two ODEs for the fluid variables $W'(r)$ and $\delta\rho'(r)$, while from $T^{\mu\theta}{_{;\mu}}=0$ we derive an algebraic relation for the fluid velocity component $V$.  This system can be further simplified by using the $\theta$-$\phi$ component of the field's equations to eliminate $H_2$ and its derivative. After some manipulations we obtain a set of five coupled, first order, inhomogeneous ODEs for $\vec{\phi}=(H_1,H_0,K,W,\delta\rho)$, which can be recast in a matrix form
\begin{equation}
\frac{d\vec{\psi}}{dr}-\hat{\boldsymbol{\alpha}}\vec{\psi}=\vec{S}_2\ ,
\end{equation}
with $\vec{\psi}=(H_1,H_0,K,W,\delta\rho)$ and the non-zero coefficients of the matrix $\hat{\boldsymbol{\alpha}}$ being

\vspace{20mm}

\begin{align}
\alpha_{11}&=\frac{r b'-b+1}{2 r b}\ \ ,\ 
\alpha_{12}=\frac{ia \left(rb'+b-1\right)-r^2 \omega ^2}{r^2 \omega  b} \, , \nonumber\\
\alpha_{13}&=\frac{i \omega}{b}\ \ , \ \alpha_{15}= -\frac{16 i \pi  c_s^2 a}{\omega  b} \, , \nonumber\\
\alpha_{21}&=-\left[\frac{i (n+1)}{r^2 \omega}-\frac{i \omega }{a}\right]\ \ ,\ 
\alpha_{22}=-\frac{2 b-1}{r b}\nonumber\\
\alpha_{23}&=-\frac{(1-3 b)}{2 rb}\ \ , \ 
\alpha_{24}=-\frac{8 i a}{3 \omega  b^2+\omega  b} \, , \nonumber\\
\alpha_{31}&=-\frac{i (n+1)}{r^2 \omega}\ \ ,\ \alpha_{32}=-\frac{1}{r}\ ,\nonumber\\
\alpha_{33}&=-\frac{(1-3 b)}{2 rb}\ \ , \
\alpha_{34}=-\frac{8 i a}{3 \omega  b^2+\omega  b}\  \, , \nonumber\\
\alpha_{43}&=\frac{i \omega  (3 b+1)^2 \left(r b'+b-1\right)}{32 r^2 b a} \, , \nonumber\\
\alpha_{42}&=\frac{i (3 b+1)\left(r b'+b-1\right) \left[2 (n+1) a+r^2 \omega ^2\right]}{16 r^4 \omega  a} \, , \nonumber\\
\alpha_{44}&=\left[\frac{(9 b+1) b'}{3 b+1}-\frac{b+3}{r}\right]\frac{W}{2b} \, , \nonumber\\
\alpha_{45}&=-\frac{i \pi  (3 b+1)  \, , 
    \left[r^2 \omega ^2-2 c_s^2 (n+1)a\right]}{r^2 \omega  a} \, , \nonumber\\
\alpha_{51}&=-\frac{\left(r b'+b-1\right) \left[(n+1) (b+1) a+2 r^2 \omega ^2 b\right]}{i32 \pi 
   c_{sr}^2 r^4 \omega  b a} \, , \nonumber\\
\alpha_{52}&=\frac{(3 b-1) \left(rb'+b-1\right)}{32 \pi  c_{sr}^2 r^3 b} \, , \nonumber\\
\alpha_{53}&=-\frac{(b+1) (3 b-1)\left(r b'+b-1\right)}{64 \pi  c_{sr}^2 r^3 b^2} \, , \nonumber\\
\alpha_{54}&=-\frac{r (b+1) a b'+b^2 a-4 r^2 \omega ^2 b-a}{i 4 \pi  c_{sr}^2 r^2 \omega  b^2 (3 b+1)} \, , \nonumber\\
\alpha_{55}&=\frac{b \left(-4 c_{s}^2+3 c_{sr}^2-1\right)+c_{sr}^2+1}{2 c_{sr}^2 r b} \, .
\end{align}
The particle contributions enter as a source term for the metric variables
\begin{equation}
\vec{S}=({S_1,S_2,S_3,0,0})\delta (\omega-m\Omega_p)\delta (r-r_p) P^{\ell}_{m}\ ,
\end{equation}
where
\begin{align}
S_{1}&=-im_p \frac{C\sqrt{\pi}L_p}{(n+1)r^4}\bigg[4r^2m-L_p(1+n-m^2) \nonumber\\
&\frac{a+b[(b-2)a-4r^2\omega^2]+r(b-1)ab'}{E_pn\omega b}\bigg]
\ ,\\
S_{2}&=-m_p C\frac{2 \sqrt{\pi}  L_p^2 (b-1) \left(n+1-m^2\right)}{ E_p n (n+1) r^3 }\ ,\\
S_{3}&=-m_p C\frac{4 \sqrt{\pi } L_p^2 b \left(n+1-m^2\right)}{ E_p n (n+1) r^3}\ ,
\end{align}
with 
\begin{equation}
C=\sqrt{\frac{(2 \ell +1) a (\ell-m)!}{b(\ell+m)!}}\ .
\end{equation}
In the previous expressions 
$P^{\ell}_{m}=P^{\ell}_{m}[cos(\theta_{p})]$ are the associate Legendre polynomials evaluated at $\theta_p=\pi/2$.

\subsection{Vacuum limit and gauge invariance}

In vacuum, when $M_\text{halo} = 0$ and $a(r) = b(r) = 1-2M_\text{BH}/r$, the axial sector master equation reduces to the familiar Regge-Wheeler equation \cite{Martel:2003jj}. 

For the polar sector, if we set  $M_\text{halo} = 0$, the equation for $\delta \rho$ decouples from the metric perturbations and becomes sourceless. This means if no initial data is given to $\delta \rho$, it remains $0$ for the whole evolution as it is the case for an EMRI in vacuum. Moreover, in vacuum, Eqs. \eqref{eq:WaveEquationK} and~\eqref{eq:WaveEquationS}, together with the constraint~\eqref{eq:HamiltonianConstraint}, can be reduced to the single Zerilli master equation \cite{Zerilli:1970wzz} for the variable~\cite{Martel:2003jj}
\beq
Z_{lm} = \frac{r}{n+1}\left[K + \frac{a}{n}\left(H_2 - r \frac{\partial K}{\partial r} \right) \right] \, .
\eeq

These master functions are gauge invariant and are thus useful to compute quantities of interest such as the flux of energy carried by GWs to infinity which is given by
\beq
\dot{E}_\text{GW}^{\infty}=\lim_{r\rightarrow \infty}\frac{1}{32\pi}\sum_{\ell=2}^{\infty}\sum_{m=-\ell}^\ell 
\frac{(\ell+2)!}{(\ell-2)!}\left(\vert Z_{\ell m} \vert^2+
4\vert \chi_{\ell m}\vert^2\right)\ . \nonumber \\
\label{math:totalflux}
\eeq

In this work, we did not explore if a gauge-invariant formulation of the perturbation equations is possible in non-vacuum background spacetimes. However, since the background spacetime is asymptotically flat, we can still make use of these properties in the limit $r \rightarrow \infty$. In practice, we achieve this numerically by extracting the value of the fields at sufficiently large radius such that $\rho^{(0)}$ and $\delta \rho$ are negligible for the intended precision of our results.

\bibliography{biblio}

\end{document}